\documentclass[graybox]{svmult}

\usepackage{mathptmx}       % selects Times Roman as basic font
\usepackage{helvet}         % selects Helvetica as sans-serif font
\usepackage{courier}        % selects Courier as typewriter font
\usepackage{makeidx}         % allows index generation
\usepackage{graphicx}        % standard LaTeX graphics tool
\usepackage{multicol}        % used for the two-column index
\usepackage[bottom]{footmisc}% places footnotes at page bottom
\usepackage{chicago}
\usepackage{amsmath,amssymb}

\makeindex             % used for the subject index
\newcommand{\be}{\begin{eqnarray}} 
\newcommand{\ee}{\end{eqnarray}} 
\newcommand{\bea}{\begin{eqnarray*}} 
\newcommand{\eea}{\end{eqnarray*}}

\begin{document} 
\bibliographystyle{chicago}

\title*{The role of the nugget term in the Gaussian process method}
\titlerunning{The role of the nugget term in the Gaussian process method} %for an abbreviated version of
\author{Andrey Pepelyshev}
\institute{Dr A. Pepelyshev \at University of Sheffield, Sheffield, S3 7RH, UK,
\email{a.pepelyshev@sheffield.ac.uk} }
\maketitle

\abstract{
The maximum likelihood estimate of the correlation parameter of a Gaussian process 
with and without of a nugget term is studied in the case of the analysis of deterministic models.
}

\section{Introduction} 

The Gaussian process method is an elegant way to analyze the results of experiments 
in many areas of science including machine learning (\citeNP{Rasmussen2006}), 
spatial statistics 
(\citeNP{matheron1973}, \citeNP{ripley1981}, \citeNP{cressie1993}, \citeNP{muller2007}), 
and the Bayesian analysis of computer experiments 
(\citeNP{sacks1989}, \citeNP{Kennedy2001425}, \citeNP{santner2003}).
Each area has its own specific ways of employing and interpreting the Gaussian processes.
The purpose of this paper is not to give a full overview, that can be found in the above references,
but to discuss some issues on the nugget term for the analysis of computer experiments.

The conception of the nugget term was first introduced in geostatistics
by \citeN{Matheron196233}. 
Roughly speaking, the variogram and covariance often show a discontinuity 
at the origin, termed the  nugget effect. 
The nugget effect is considered as a random noise and may represent a measurement error
or short scale variabilities.
The nugget term is a well explored object in spatial statistics 
(\citeNP{Pitard1993}).

Another area of the application of Gaussian processes is 
the Bayesian approach developed for the analysis of computer experiments. 
In this approach, a so-called emulator is introduced for making
probabilistic judgments on the true output of the given computer model, which is called a simulator.
A Gaussian process is used for a full probabilistic specification of the emulator.
Thus, the emulator is utilized to measure uncertainty of different kinds, see (\citeNP{Kennedy2001425}).

Formally, there is no nugget term in the Gaussian process method for the analysis of deterministic models,
but the nugget term can be introduced artificially, for example,
for the regularization of the inversion of a covariance matrix, see  (\citeNP{Neal1997}) for details.
\citeN{Gramacy2009130} reported on the usefulness of the nugget term
in their research of supercomputer experiments.

The presence of the nugget term in the Gaussian process method is natural 
for the analysis of stochastic and simulation models.
The nugget effect may represent a measurement error or an effect of random values used inside computer models 
(\citeNP{kleijnen2008}, \citeNP{kleijnen2005}).

The influence of the nugget term for optimal designs of experiments
for a number of cases have been studied in
(\citeNP{zhu2005}, \citeNP{stehlik2008}).

The present paper focuses on the Gaussian process method applied for the analysis of deterministic models.
It is shown that the nugget term has a great impact on the likelihood and the estimate of correlation parameter.

%\newpage 
\section{The likelihood for a Gaussian process without the nugget term} 

In this section, it is shown that the likelihood of a Gaussian process has an unexpected behaviour in 
the analysis of non-stochastic models.
More precisely, for a deterministic model of observations,
the maximum likelihood estimate of the correlation parameter 
may tend to the infinity as the number of points increases.
It means that a deterministic model is approximated by a Gaussian process with the correlation function 
$r(x)\approx1$ for any $x$.
%To the best of our knowledge, this fact is not presented in the literature.

Indeed, let $y_i=\eta(x_i)$ be the output of the model $\eta(x)$ at
the point $x_i\in[0,1]$, $i=1,\ldots,n$. Note that
for a deterministic model, the replication of an observation at some point gives the same output.
Without loss of generality, let $x_1<\ldots<x_n$.
The likelihood for a Gaussian process with constant mean $\beta$, 
variance $\sigma^2$ and correlation function $r(x,\tilde x)=e^{-|x-\tilde x|/\psi}$ have the form
\bea
 p(y|\beta,\sigma,\psi)=\frac{|R|^{-1/2}}{(2\pi\sigma^2)^{n/2}}
 e^{-\frac{1}{2\sigma^2}(y-H\beta)^TR^{-1}(y-H\beta)}
\eea
where $y=(y_1,\ldots,y_n)^T$ is the vector of output values, 
$R=(r(x_i,x_j|\psi))_{i,j=1}^n$ is the correlation matrix, 
$H=(h(x_1),\ldots,h(x_n))$, and $h(x)\equiv1$.

\begin{figure}[!hhh]
\centering
\includegraphics[width=50mm]{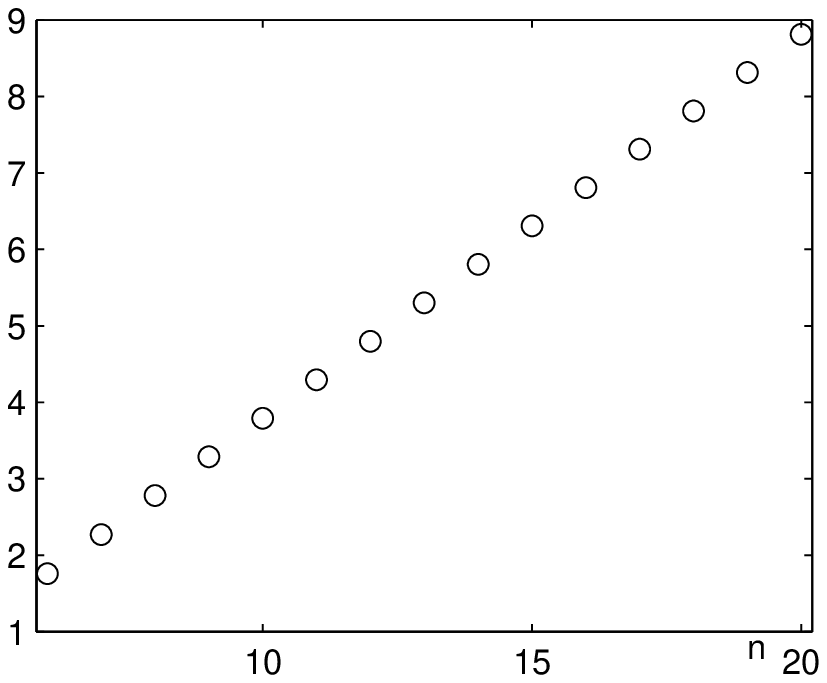}
\includegraphics[width=50mm]{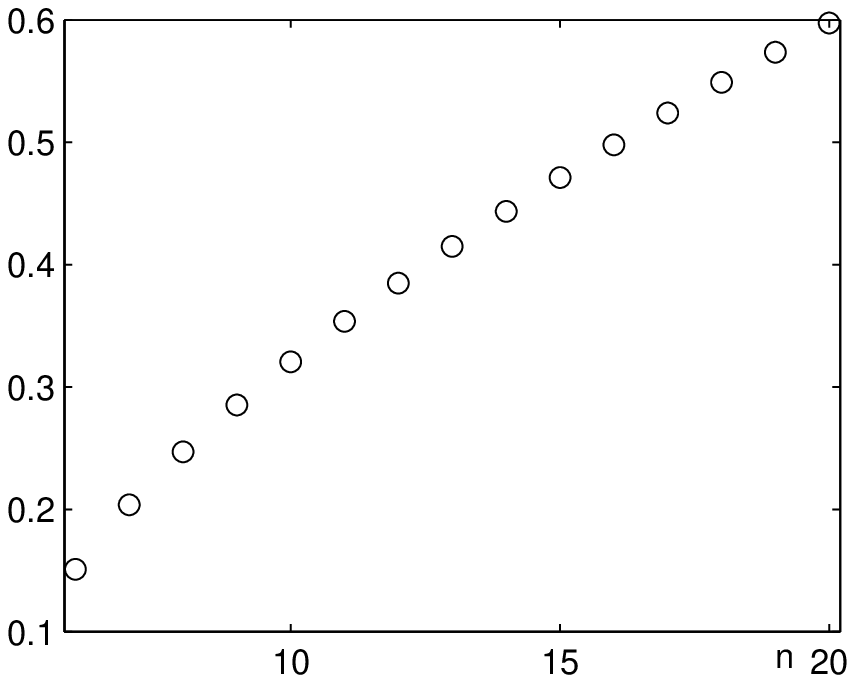}\kern-15mm\\
\caption{The maximum likelihood estimate of $\psi$ for the Gaussian process  
with the exponential correlation function and $n$ equidistant points on the interval $[0,1]$
for the model $\eta(x)=x-1/2$ (left part) and 
for the model $\eta(x)=\sin(2\pi x)$ (right part) for $n=6,\ldots,20$.}
\label{fig:ml-exp}       % Give a unique label
\end{figure}

The maximum likelihood (ML) estimates of $\beta$ and $\sigma$ have the following explicit forms
\bea
 \hat\beta_{ML}=(H^TR^{-1}H)^{-1}HR^{-1}y
\eea
and
\bea
 \hat\sigma^2_{ML}=\frac{1}{n}(y-H\hat\beta_{ML})^TR^{-1}(y-H\hat\beta_{ML}).
\eea
The ML estimate of $\psi$ can be found only numerically
in the following way
\bea
 \hat\psi_{ML}=\arg\max_{\psi\in(0,\infty)} p(y|\hat\beta_{ML},\hat\sigma_{ML},\psi).
\eea\vskip-1mm\noindent
After substituting and simplifying, we obtain that the estimate $\hat\psi_{ML}$ maximizes
\vspace{-1mm}
\bea
 L(\psi)=\ln\left[|R|^{-1/2}\right]-
 \frac{n}{2}\ln\left[(y-H\hat\beta_{ML})^TR^{-1}(y-H\hat\beta_{ML})\right].
\eea\vskip-1mm\noindent
For the exponential correlation function, the inverse of matrix $R$ admits 
the explicit representation $R^{-1}=V^TV$
where the matrix $V$ is defined by
\vspace{-1mm}
{%\small%\footnotesize
\bea
V  = \left[ {\begin{array}{*{20}c}
   1 & 0 & 0 &  \cdots  & 0 & 0  \\
   { - \frac{{\mu_2 }}{{\sqrt {1 - \mu_2^2 } }}}
   & {\frac{1}{{\sqrt {1 - \mu_2^2 } }}}
   & 0 & \cdots  & 0 & 0  \\
   0
   & { - \frac{{\mu_3 }}{{\sqrt {1 - \mu_3^2 } }}}
   & {\frac{1}{{\sqrt {1 - \mu_3^2 } }}} & \cdots  & 0 & 0  \\
   &   &   &  &  &    \\
   \vdots  &  \vdots  & \vdots  & \ddots  & \ddots &  \vdots   \\
   0 & 0 & 0 &  \cdots
   & { - \frac{{\mu_n }}{{\sqrt {1 - \mu_n^2 } }}}
   & {\frac{1}{{\sqrt {1 - \mu_n^2 } }}}  \\
\end{array}} \right],
\eea}\vskip-1mm\noindent
$\mu_i=e ^{-(x_i  - x_{i-1} )/\psi}$.
For $n$ equidistant points $x_i=(i-1)/(n-1)$, $i=1,\ldots,n$, straightforward calculation shows that 
\vspace{-1mm}
\bea
 yR^{-1}y=\frac{y_1^2+y_n^2}{ {1 - \lambda^2}}+
 \sum_{i=2}^{n-1}y_i^2\frac{1 + \lambda^2}{{1 - \lambda^2}}
-2\sum_{i=1}^{n-1} y_iy_{i+1}\frac{ \lambda}{{1 - \lambda^2}}
\eea\vskip-1mm\noindent
where $\lambda=e ^{-\frac{1}{(n-1)\psi}}$, and 
\vspace{-1mm}
\bea
 |R|^{-1/2}=\frac{1}{ ({1 - \lambda^2})^{(n-1)/2}}\,.
\eea\vskip-1mm\noindent
For the model $\eta(x)=x-1/2$, we obtain that $\hat\beta_{ML}=0$ and
\vspace{-1mm}
\bea
 yR^{-1}y=\frac{1}{2}\frac{1}{ {1 - \lambda^2}}+
 \frac{n^2-5n+6}{12(n-1)}\cdot\frac{1 + \lambda^2}{{1 - \lambda^2}}
-\frac{n^2-2n-3}{6(n-1)}\cdot\frac{ \lambda}{{1 - \lambda^2}}\,.
\eea\vskip-1mm\noindent
The estimate $\hat\psi_{ML}$ can be found explicitly in Maple and 
is not presented since it is a very large expression. Applying the power series expansion, we have
\vspace{-1mm}
\bea
 e ^{-\frac{1}{(n-1)\hat\psi_{ML}}}\!=\!1-\frac{2}{n^2}-\frac{20}{3n^2}+O\left(\frac{1}{n^3}\!\right)
 \mbox{ and }\hat\psi_{ML}\!=\!\frac{n}{2}-\frac{7}{6}-\frac{7}{18n}-\frac{17}{54n^2}+O\left(\frac{1}{n^3}\!\right).
\eea\vskip-1mm\noindent
The dependence of $\hat\psi_{ML}$ on $n$  is given 
in Figure \ref{fig:ml-exp} for the model $\eta(x)=x-1/2$ at the left part and 
for the model $\eta(x)=\sin(2\pi x)$ at the right part.
We observe that the estimate $\hat\psi_{ML}$ increases almost linearly as $n$ increases 
for both models.

\begin{figure}[!hhh]
\centering
\includegraphics[width=50mm]{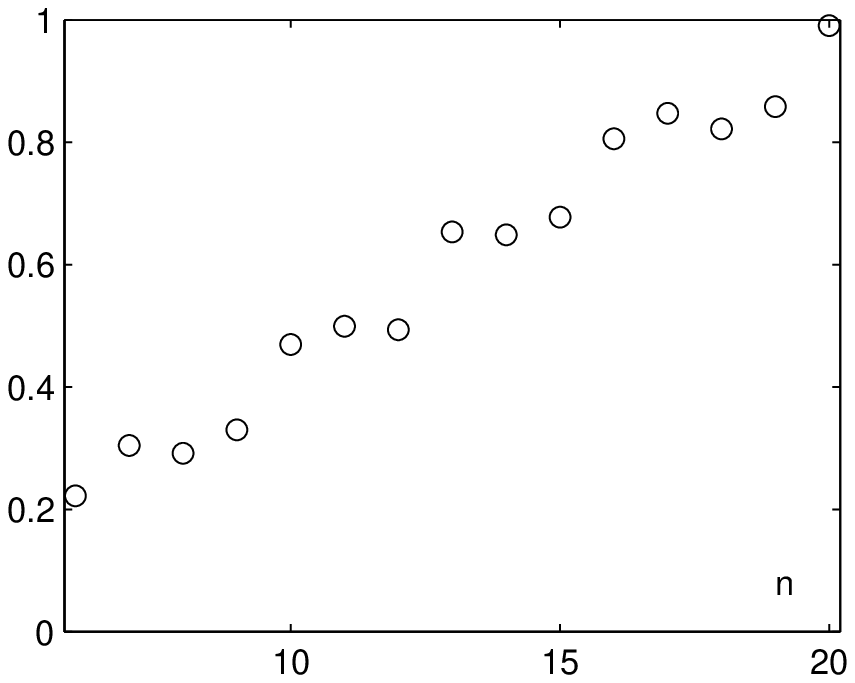}
\includegraphics[width=50mm]{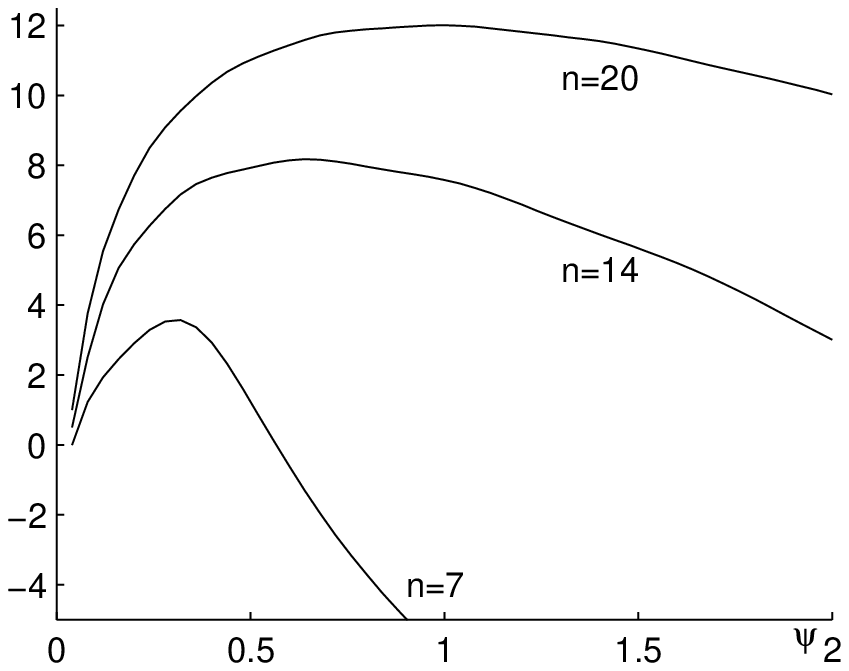}\kern-15mm\\
\caption{At left: The maximum likelihood estimate of $\psi$ for the Gaussian process 
with the Gaussian correlation function and $n$ equidistant points on the interval $[0,1]$
for the model $\eta(x)=\sin(2\pi x)$ for $n=6,\ldots,20$.
At right: The likelihood function of $\psi$ for $n=7,14,20$.}
\label{fig:ml-gaus}       % Give a unique label
\end{figure}

The maximum likelihood estimate of $\psi$ for the Gaussian correlation function
$r(x,\tilde x)=e^{-(x-\tilde x)^2/\psi}$ is given in Figure \ref{fig:ml-gaus}.
For the model $\eta(x)=x-1/2$ we have that $\hat\psi_{ML}=\infty$ for any $n$. 
Note that for the model $\eta(x)=\sin(2\pi x)$, 
the condition number of the correlation matrix $R(\hat\psi_{ML})$
is of order $10^7$, $10^{14}$, $10^{22}$, $10^{30}$, and $10^{38}$ for $n=8,$ $11,$ $14,$ $17,$ and $20$,
respectively. These calculations were done in Maple with 45 digits precision.
However, the computer representation of floating numbers typically has only 17 digits.
Thus, it is impossible to find the maximum likelihood estimate for large $n$
using the ordinary floating representation in a computer.
In particular, \citeN{Ababou199499} have shown that  
the condition number grows linearly for the exponential correlation function
and grows exponentially for the Gaussian correlation function.

%\begin{figure}[!hhh]
%\centering
%\includegraphics[width=50mm]{lik_gsin.eps}\kern-15mm\\
%\caption{The likelihood for $\psi$ using Gaussian process method 
%with Gaussian correlation function and $n$ equidistant points on the interval $[0,1]$
%for the model $\eta(x)=\sin(2\pi x)$ for $n=7,14,20$.}
%\label{fig:lik-gaus}       % Give a unique label
%\end{figure}

In more general situations for other correlation functions and other models,
the dependence of the maximum likelihood estimate and the restricted maximum likelihood estimate of 
$\psi$ on $n$ remains typically the same and can be verified numerically (\citeNP{Pepelyshev}).

Thus, roughly speaking, the estimate of parameters of a Gaussian process is associated with the given data set 
and is not associated with the deterministic model.
This estimation is not simple and is not well-defined. 
It is easy to observe that if one divides an input space into several regions, 
one may get quite different estimates of parameters for different regions. 
However, if one is looking for one Gaussian process 
over the full space, one has difficulty in finding the single estimate.

\section{The likelihood for a Gaussian process with the nugget term}
\subsection{MLE for a Gaussian process}

In this section, the likelihood with the presence of the nugget term is investigated.
For this case, the correlation matrix $R$ in
formulae from Section 2 should be replaced to the correlation matrix
\vspace{-1mm}
\bea
 R_\nu=\big((1-\nu)r(x_i-x_j)+\nu\delta_{i,j}\big)_{i,j}
\eea\vskip-1mm\noindent
where $\nu$ is the nugget term.

\vspace{-2mm}
\begin{figure}[!hhh]
\centering
\includegraphics[width=50mm]{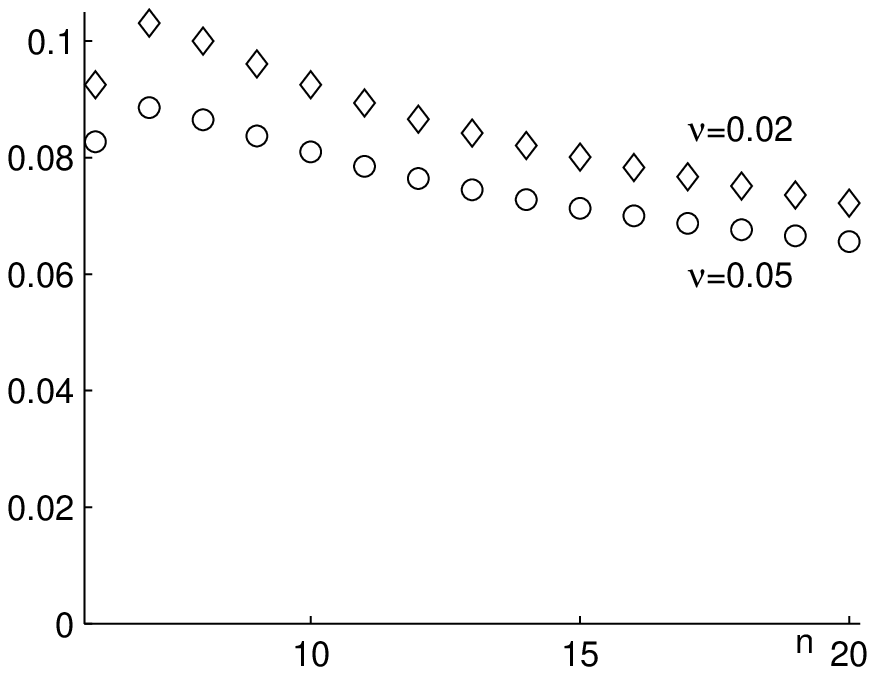}
\includegraphics[width=50mm]{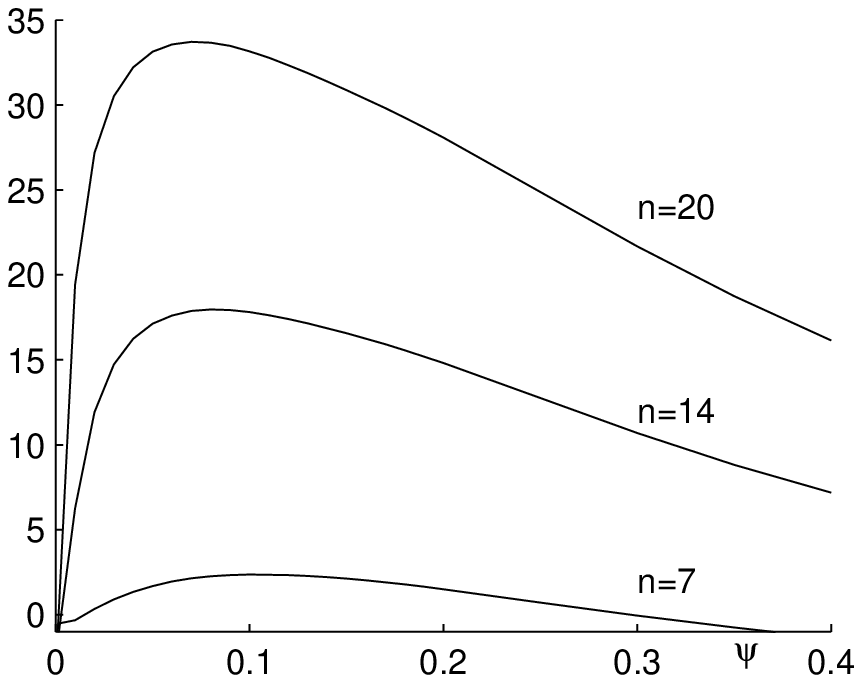}\kern-15mm\\
\caption{At left: The maximum likelihood estimate of $\psi$ for the Gaussian process  
with the Gaussian correlation function and the nugget term $\nu=0.02,0.05$ 
for measurements of  the model $\eta(x)=\sin(2\pi x)$ at $n$ equidistant points on the interval $[0,1]$,
$n=6,\ldots,20$.
At right: The likelihood function of $\psi$ for the nugget term $\nu=0.02$ and 
for $n=7,14,20$.}
\label{fig:ml-gaus_nug}       % Give a unique label
\end{figure}

The likelihood function and the maximum likelihood estimate for fixed values of the nugget term
are presented in Figure \ref{fig:ml-gaus_nug}.
One can observe that the nugget term essentially changes the maximum likelihood estimate of $\psi$
(and also $\sigma$). 
The estimate $\hat\psi_{ML}$ does not increase to infinity as $n$ increases, since
the Gaussian process is fitted to a band around the deterministic function.
It should also be noted that the condition number of the correlation matrix $R_\alpha$ is of order $10^2$
and is increasing very slowly as $n$ is increasing.
Moreover, the estimate $\hat\psi_{ML}$ is smaller with the presence of the nugget term
that also reduces the condition number of the correlation matrix.
\citeN{Ababou199499} have shown that the condition number of the correlation matrix 
for the Gaussian process models increases to a finite limit with the presence of the nugget term.

\begin{figure}[!hhh]
\centering
\includegraphics[width=100mm]{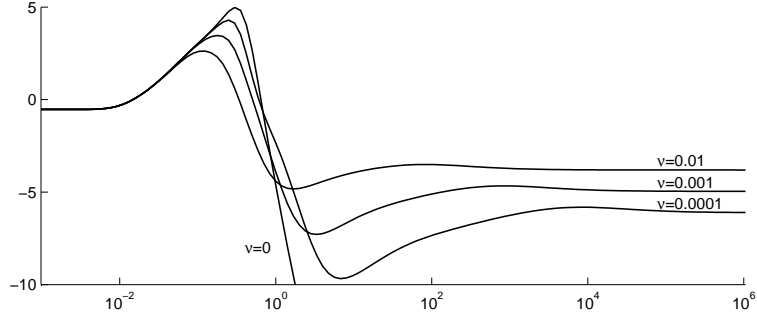}
\caption{The likelihood function of $\psi$ for the Gaussian process  
with the Gaussian correlation function and the nugget term $\nu=0,0.01,0.001,0.0001$ 
for measurements of the model $\eta(x)=\sin(2\pi x)$ at $7$ equidistant points on the interval $[0,1]$.}
\label{fig:liknu}       % Give a unique label
\end{figure}

Note one undesired effect of the nugget term. 
The likelihood may have the second mode for large values of the correlation parameter,
see Figure~\ref{fig:liknu}.
The second mode strongly depends on a value of the nugget term and can be considered as a false mode.
For some data, the likelihood function at the second mode may have a larger value than at the first mode.

Note that in the presence of the nugget term, the meta-model
\vspace{-0mm}
\bea
 m_\nu(x)=H\beta+t^T(x)R^{-1}_\nu(y-H\beta)
\eea\vskip-0mm\noindent
where $t(x)=(r(x,x_1),\ldots,r(x,x_n))^T$ does not possess the interpolation  property.
Nevertheless, the deviations $\varepsilon_i=y_i-m_\nu(x_i)$ are very small.
One may construct a meta-model, that interpolates the dataset $\{(x_i,\varepsilon_i)\}_{i=1}^n$,
by a method given in Cressie (1993, Sect. 5.9).
It is not necessary for the deviations $\varepsilon_i$ to use the Kriging approach without the nugget term.
One may use the inverse distance weighted interpolation (\citeNP{cressie1993}, p. 371, \citeNP{Lu20081044})
and define the meta-model in the following form
\vspace{-0mm}
\bea
 m(x)=m_\nu(x)+\frac{\sum\limits_{i=1}^n \varepsilon_i {||x-x_i||_2^{-2}}}
 {\sum\limits_{i=1}^n {||x-x_i||_2^{-2}}}\,.
\eea\vskip-0mm\noindent

\subsection{MLE for stationary processes}

Let us perform a small simulation study.
Assume that the results of experiments satisfy
\bea
 y(x_i)=\beta+\sigma^2\varepsilon^{(1)}(x_i)+\tau^2\varepsilon^{(2)}(x_i)
\eea
where $x_1,\ldots,x_n$ are points of measurements, $\varepsilon^{(1)}(x)$ denotes a
stationary Gaussian process with correlation function $r(x)=e^{-x^2/\psi}$ and $\varepsilon^{(2)}(x)$ is white noise. 
Let $\mathbf{E}\varepsilon^{(j)}(x)=0$, $\mathbf{D}\varepsilon^{(j)}(x)=1$,
processes $\varepsilon^{(1)}(x)$ and $\varepsilon^{(2)}(x)$ be independent.
The values $\beta+\sigma^2\varepsilon^{(1)}(x_i)$ may be conceived as true values of a physical process.
The values $\tau^2\varepsilon^{(2)}(x_i)$ may be interpreted as a measurement error or a rough rounding
of measured values.
Let us compute the maximum likelihood estimators for 1000 realizations obtained for $n=8$, 
$x_i=(i-1)/7$, $i=1,\ldots,8$, $\beta=2$, $\psi=1.5$, $\sigma=1$, $\tau=0$ or $\tau=0.01$.
Results of the maximum likelihood estimation for different values of the nugget term are presented in Table~\ref{P:mle}.

\begin{table}[!hhh]
  \caption{The mean of maximum likelihood estimators of parameters using different values of the nugget term. 
  Standard deviations are given in brackets.}
  \label{P:mle}
  \smallskip
  \centering
{\small
  \begin{tabular}{|c| c|c|c| c|c|c|}
  \hline
  &\multicolumn{3}{|c|}{$\tau=0$}&\multicolumn{3}{|c|}{$\tau=0.01$}\\
  \hline
  $\nu$&0&0.01&0.02&0&0.01&0.02\\
  \hline
  $\hat\beta_{ML}$&  ~2.03(0.68)&~2.01(0.85)&~2.02(0.86)& ~2.02(0.92)&~2.04(0.85)&~2.04(0.86)\\
  $\hat\sigma_{ML}$& ~0.83(0.40)&~0.29(0.17)&~0.27(0.16)& ~0.33(0.23)&~0.30(0.17)&~0.28(0.16)\\
  $\hat\psi_{ML}$&   ~1.44(0.37)~&~0.54(0.25)~&~0.47(0.20)~& ~0.14(0.06)~&~0.58(0.29)~&~0.49(0.23)~\\
  \hline
  \end{tabular}}
\end{table}

\noindent
One can observe that the maximum likelihood estimators with a nonzero nugget term 
does not depend on small perturbations $\{\tau^2\varepsilon^{(2)}(x_i)\}_i$ of the data
$\{\beta+\sigma^2\varepsilon^{(1)}(x_i)\}_i$.
In contrast, for $\nu=0$, the maximum likelihood estimators of $\sigma$ and $\psi$ are significantly changed
due to adding small perturbations.
In all cases, the accuracy of $\hat\beta_{ML}$ is approximately the same.
Thus, as can be seen, the nugget term yields a regularization effect on the maximum likelihood estimators.

\section{Conclusions}

In the analysis of deterministic models
the presence of a nugget term has a significant impact on the likelihood of a Gaussian process.
The maximum likelihood estimate of the correlation parameter with a nonzero nugget term is more reliable
and the condition number of the correlation matrix is moderate.
Even if a deterministic model does not have any internal computational errors 
or other perturbations,
the artificial introduction of the nugget term can be recommended.

\begin{acknowledgement}
Andrey Pepelyshev thanks two referees for their valuable comments and suggestions, 
and acknowledges the financial support provided
by the MUCM project (EPSRC grant EP/D048893/1, http://mucm.group.shef.ac.uk).
\end{acknowledgement}

%\bibliography{pepelyshev}

\begin{thebibliography}{19.}

\bibitem[\protect\citeauthoryear{Ababou, Bagtzoglou, and Wood}{Ababou
  et~al.}{1994}]{Ababou199499}
Ababou, R., A.~Bagtzoglou, and E.~Wood (1994).
\newblock On the condition number of covariance matrices in kriging,
  estimation, and simulation of random fields.
\newblock {\em Mathematical Geology\/}~{\em 26\/}(1), 99--133.

\bibitem[\protect\citeauthoryear{Cressie}{Cressie}{1993}]{cressie1993}
Cressie, N. A.~C. (1993).
\newblock {\em Statistics for spatial data}.
\newblock Wiley Series in Probability and Mathematical Statistics: Applied
  Probability and Statistics. New York: John Wiley \& Sons Inc.
\newblock Revised reprint of the 1991 edition, A Wiley-Interscience
  Publication.

\bibitem[\protect\citeauthoryear{Gramacy and Lee}{Gramacy and
  Lee}{2009}]{Gramacy2009130}
Gramacy, R. and H.~Lee (2009).
\newblock Adaptive design and analysis of supercomputer experiments.
\newblock {\em Technometrics\/}~{\em 51\/}(2), 130--145.

\bibitem[\protect\citeauthoryear{Kennedy and O'Hagan}{Kennedy and
  O'Hagan}{2001}]{Kennedy2001425}
Kennedy, M.~C. and A.~O'Hagan (2001).
\newblock Bayesian calibration of computer models.
\newblock {\em Journal of the Royal Statistical Society. Series B\/}~{\em
  63\/}(3), 425--450.

\bibitem[\protect\citeauthoryear{Kleijnen}{Kleijnen}{2008}]{kleijnen2008}
Kleijnen, J. P.~C. (2008).
\newblock {\em Design and analysis of simulation experiments}.
\newblock International Series in Oper. Res. \& Management Science, 111. New
  York: Springer.

\bibitem[\protect\citeauthoryear{Kleijnen and van Beers}{Kleijnen and van
  Beers}{2005}]{kleijnen2005}
Kleijnen, J. P.~C. and W.~C.~M. van Beers (2005).
\newblock Robustness of {K}riging when interpolating in random simulation with
  heterogeneous variances: some experiments.
\newblock {\em European J. Oper. Res.\/}~{\em 165\/}(3), 826--834.

\bibitem[\protect\citeauthoryear{Lu and Wong}{Lu and Wong}{2008}]{Lu20081044}
Lu, G. and D.~Wong (2008).
\newblock An adaptive inverse-distance weighting spatial interpolation
  technique.
\newblock {\em Computers and Geosciences\/}~{\em 34\/}(9), 1044--1055.

\bibitem[\protect\citeauthoryear{Matheron}{Matheron}{1962}]{Matheron196233}
Matheron, G. (1962).
\newblock Traite de geostatistique appliquee. memoires bur rech.
\newblock {\em Geol Minieres\/}~{\em 24}.

\bibitem[\protect\citeauthoryear{Matheron}{Matheron}{1973}]{matheron1973}
Matheron, G. (1973).
\newblock The intrinsic random functions and their applications.
\newblock {\em Advances in Appl. Probability\/}~{\em 5}, 439--468.

\bibitem[\protect\citeauthoryear{M{\"u}ller}{M{\"u}ller}{2007}]{muller2007}
M{\"u}ller, W.~G. (2007).
\newblock {\em Collecting spatial data\/} (revised ed.).
\newblock Contributions to Statistics. Heidelberg: Physica-Verlag.
\newblock Optimum design of experiments for random fields.

\bibitem[\protect\citeauthoryear{Neal}{Neal}{1997}]{Neal1997}
Neal, R. (1997).
\newblock Monte carlo implementation of gaussian process models for bayesian
  classification and regression.
\newblock {\em Technical Report\/}~{\em 9702}.

\bibitem[\protect\citeauthoryear{Pepelyshev}{Pepelyshev}{2009}]{Pepelyshev}
Pepelyshev, A. (2009).
\newblock Fixed-domain asymptotics of maximum likelihood estimators for
  observations of deterministic models.
\newblock {\em submitted\/}.

\bibitem[\protect\citeauthoryear{Pitard}{Pitard}{1993}]{Pitard1993}
Pitard, F. (1993).
\newblock {\em Exploration of the Nugget Effect}.
\newblock Doordrecht, The Netherlands: Kluwer Academic Pub.

\bibitem[\protect\citeauthoryear{Rasmussen and Williams}{Rasmussen and
  Williams}{2006}]{Rasmussen2006}
Rasmussen, C.~E. and C.~K.~I. Williams (2006).
\newblock {\em Gaussian processes for machine learning}.
\newblock Adaptive Comp. and Machine Learning. Cambridge: MIT Press.

\bibitem[\protect\citeauthoryear{Ripley}{Ripley}{1981}]{ripley1981}
Ripley, B.~D. (1981).
\newblock {\em Spatial statistics}.
\newblock New York: John Wiley \& Sons Inc.
\newblock Wiley Series in Probability and Mathematical Statistics.

\bibitem[\protect\citeauthoryear{Sacks, Welch, Mitchell, and Wynn}{Sacks
  et~al.}{1989}]{sacks1989}
Sacks, J., W.~J. Welch, T.~J. Mitchell, and H.~P. Wynn (1989).
\newblock Design and analysis of computer experiments.
\newblock {\em Statist. Sci.\/}~{\em 4\/}(4), 409--435.
\newblock With comments and a rejoinder by the authors.

\bibitem[\protect\citeauthoryear{Santner, Williams, and Notz}{Santner
  et~al.}{2003}]{santner2003}
Santner, T.~J., B.~J. Williams, and W.~I. Notz (2003).
\newblock {\em The design and analysis of computer experiments}.
\newblock Springer Series in Statistics. New York: Springer-Verlag.

\bibitem[\protect\citeauthoryear{Stehl{\'{\i}}k, Rodr{\'{\i}}guez-D{\'{\i}}az,
  M{\"u}ller, and L{\'o}pez-Fidalgo}{Stehl{\'{\i}}k et~al.}{2008}]{stehlik2008}
Stehl{\'{\i}}k, M., J.~M. Rodr{\'{\i}}guez-D{\'{\i}}az, W.~G. M{\"u}ller, and
  J.~L{\'o}pez-Fidalgo (2008).
\newblock Optimal allocation of bioassays in the case of parametrized
  covariance functions: an application to lung's retention of radioactive
  particles.
\newblock {\em TEST\/}~{\em 17\/}(1), 56--68.

\bibitem[\protect\citeauthoryear{Zhu and Stein}{Zhu and Stein}{2005}]{zhu2005}
Zhu, Z. and M.~L. Stein (2005).
\newblock Spatial sampling design for parameter estimation of the covariance
  function.
\newblock {\em J. Statist. Plann. Inference\/}~{\em 134\/}(2), 583--603.

\end{thebibliography}
 
%\section*{References}

\end{document}